\documentclass[twocolumn]{aa} 

\usepackage{graphicx}
\usepackage{txfonts}
\begin{document}

\title{Recurring obscuration in NGC~3783}

\author{J.S. Kaastra\inst{1,2}
\and
M. Mehdipour\inst{1}
\and 
E. Behar\inst{3}
\and
S. Bianchi\inst{4}
\and
G. Branduardi-Raymont\inst{5}
\and
L. Brenneman\inst{6}
\and
M. Cappi\inst{7}
\and
E. Costantini\inst{1}
\and
B. De Marco\inst{8}
\and
L. di Gesu\inst{9}
\and
J. Ebrero\inst{10}
\and
G.A. Kriss\inst{11}
\and
J. Mao\inst{1,2}
\and 
U. Peretz\inst{3}
\and
P.-O. Petrucci\inst{12}
\and
G. Ponti\inst{13}
\and
D. Walton\inst{14}
}

\institute{
SRON Netherlands Institute for Space Research, Sorbonnelaan 2,
3584 CA Utrecht, the Netherlands 
\and
Leiden Observatory, Leiden University, PO Box 9513,
2300 RA Leiden, the Netherlands
\and
Department of Physics, Technion-Israel Institute of Technology, 
Haifa 32000, Israel 
\and 
Dipartimento di Matematica e Fisica, Universit\`a degli Studi Roma Tre, 
Via della Vasca Navale 84, 00146 Roma, Italy 
\and
Mullard Space Science Laboratory, University College London, Holmbury St. Mary,
Dorking, Surrey, RH5 6NT, UK
\and
Smithsonian Astrophysical Observatory, Cambridge, Massachusetts, USA
\and
INAF-IASF Bologna, via Gobetti 101, 40129, Bologna, Italy
\and
Nicolaus Copernicus Astronomical Center, Polish Academy of Sciences, Bartycka
18, 00-716, Warsaw, Poland
\and
Department of Astronomy, University of Geneva, 16 Ch. d'Ecogia, 1290, Versoix, Switzerland
\and
European Space Astronomy Centre, PO Box 78, 28691 Villanueva de la
Ca\~{n}ada, Madrid, Spain
\and
Space Telescope Science Institute, 3700 San Martin Drive, Baltimore, MD, 21218, USA
\and
Univ. Grenoble Alpes, CNRS, IPAG, 38000, Grenoble, France
\and
Max Planck Institute f\"ur Extraterrestriche Physik, 85748, Garching, Germany
\and
Institute of Astronomy, Madingley Road, CB3 0, HA Cambridge, UK
}

\date{Received \today; accepted \today}

 
  \abstract
{Obscuration of the continuum emission from active galactic nuclei by streams of
gas with relatively high velocity ($> 1000$~km\,s$^{-1}$) and column density ($>
3\times 10^{25}$~m$^{-2}$) has been seen in a few Seyfert galaxies. This
obscuration has a transient nature. In December 2016 we have witnessed such an
event in NGC~3783.}
{The frequency and duration of these obscuration events is poorly known. Here we
study archival data of NGC~3783 in order to constrain this duty cycle.}
{We use archival Chandra/NuSTAR spectra taken in August 2016. We also study the
hardness ratio of all Swift XRT spectra taken between 2008--2017.}
{In August 2016, NGC~3783 also showed evidence for obscuration. While the column
density of the obscuring material is ten times lower than in December 2016, the
opacity is still sufficient to block a significant fraction of the ionising
X-ray and EUV photons. From the Swift hardness ratio behaviour we find several
other epochs with obscuration. Obscuration with columns
$>10^{26}$~m$^{-2}$ may take place in about half of the time. Also in archival
X-ray data taken by ASCA in 1993 and 1996 we find evidence for obscuration.}
{Obscuration of the ionising photons in NGC~3783 occurs more frequently than
previously thought. This may not always have been recognised due to low spectral
resolution observations, too limited spectral bandwidth or confusion with
underlying continuum variations.}


   \maketitle
%

\section{Introduction}

Outflows from active galactic nuclei (AGN) have been found in about half of all
nearby systems \citep{crenshaw2003}. They become visible because of the imprint
of absorption lines and edges in the X-ray and UV spectra of these objects. They
show a broad range of ionisation state with column densities usually up to about
$10^{26}$~m$^{-2}$ at the highest ionisation parameters. These outflows may
originate from an accretion disk wind or torus wind, and have speeds up to a few
thousand km\,s$^{-1}$ or even much higher if ultra-fast outflows are considered.
But all these outflow components share the property that the total flux taken
away from the continuum radiation is modest.

There is also evidence for outflows with much higher column densities
and often at lower ionisation states than the ones above, which absorb
so much soft X-ray and EUV radiation that we may call them obscurers. While in
individual cases X-ray obscuration has been demonstrated to be present in CCD
spectra, the relatively low resolution of CCDs prevents investigating the
dynamics of these components. Examples for the latter are the X-ray eclipses in
Mrk~766 and NGC~1365 that were attributed to broad line region clouds passing
our line of sight \citep{risaliti2007,risaliti2011}.

This situation changed when NGC~5548 was found to be obscured in 2013
\citep{kaastra2014s}. It was observed simultaneously with high-resolution X-ray
detectors (the RGS on XMM-Newton), broad-band X-ray instruments (EPIC on
XMM-Newton, NuSTAR), high-resolution UV spectroscopy (COS on HST) and photometry
in the optical/UV band (OM on XMM-Newton, UVOT on Swift). The X-ray data
revealed two obscuring components with column densities of
$10^{26}-10^{27}$~m$^{-2}$, X-ray covering factor 0.86--0.30 and lowly ionised
to almost neutral for both components, respectively. The low ionisation
component produces broad UV absorption lines which reveal that the obscuring
material is outflowing with velocities up to 5000~km\,s$^{-1}$. In general, the
UV covering factor is about 3 times lower than the X-ray covering factor, and
the obscuring material is likely to be close to the outer broad line region.
Other sources studied with joint UV and X-ray spectroscopy, although with much
sparser observations, which show signs of obscuration are Mrk~335
\citep{longinotti2013} and NGC~985 \citep{ebrero2016}.

Triggered by these results on NGC~5548, we have started a monitoring program
with Swift on a sample of Seyfert galaxies to find them in obscured states using
the hardness ratio of the X-ray spectra as a discerning diagnostic against
simple flux variability. In an obscured state, the spectrum significantly
hardens. This program triggered successfully on NGC~3783 in December 2016
\citep{mehdipour2017}. Simultaneous observations with XMM-Newton, HST and NuSTAR
showed that NGC~3783 was in a similar obscured state as NGC~5548. However, while
the obscuration of NGC~5548 is lasting now already for several years, the event
in NGC~3783 lasted for about a month as evidenced by the Swift monitoring. 

In order to better understand the obscuration phenomenon, it is important to
know its duty cycle: how frequent are these events, and what is their duration.
NGC~3783 was selected for our monitoring program because previous monitoring
with RXTE has shown one obscuring event in March 2008 and two possible events in
July 2008 and February 2011 \citep{markowitz2014}. Swift, with its softer energy
band as compared with RXTE, is even more sensitive to detect these events, and
NGC~3783 has been monitored intensively with Swift in Spring 2017, after the
obscuration event in December 2016 described by \citet{mehdipour2017}.
Furthermore, in August 2016 coordinated Chandra HETGS and NuSTAR observations
have been performed on NGC~3783. Finally, ASCA data taken in 1993 showed
remarkable differences compared with spectra taken three years later
\citep{george1998}. In this paper we re-analyse these data in order to better
constrain the duty cycle of the obscuration in NGC~3783.

\section{Data analysis}

\subsection{Data reduction}

\begin{table*}[!htbp]
\caption{Chandra HETGS, NuSTAR and Swift UVOT spectral data used in this study}
\label{tab:obslog}
\centerline{
\begin{tabular}{lccccc}
\hline\hline
Start date and time & End date and time & ObsID & Instrument & net exposure time (ks) \\
\hline\noalign{\smallskip}
22-8-2016 07:24 & 23-8-2016 06:10 & 18192       & HETGS  & 79 \\
22-8-2016 06:26 & 23-8-2016 06:21 & 60101110002 & NuSTAR & 41 \\
22-8-2016 09:15 & 22-8-2016 09:29 & 81760001    & UVOT   &  2 \\
\hline\noalign{\smallskip}
25-8-2016 00:23 & 25-8-2016 21:30 & 19694       & HETGS  & 73 \\
24-8-2016 21:16 & 25-8-2016 21:56 & 60101110004 & NuSTAR & 42 \\
\hline\noalign{\smallskip}
\end{tabular}
}
\end{table*}

There are two Chandra/NuSTAR observations separated by three days taken in
August 2016 (PI: L Brenneman). The data used for our spectral analysis are shown
in Table~\ref{tab:obslog}.

We used the Chandra HETGS spectra from the \textit{tgcat} archive
\citep{huenemoerder2011}. The spectra were rebinned by a factor of two to
approximate the optimal binning of the data \citep{kaastra2016}. We restricted
our fits to the 2.5--26~\AA\ range for the MEG grating, and 1.55--14.5~\AA\ for
the HEG grating. We renormalise the HEG data relative to the MEG data adopting a
scaling factor of 0.954, equal to what we used for Mrk~509 \citep{kaastra2014m}.

The NuSTAR data were taken from the public archive and extracted using the
standard tools. Spectra of both detectors were combined into a single spectrum
using the \textit{HEAsoft addspec} tool. We restricted our fits to the 4--79~keV
range. We renormalised the NuSTAR data relative to the MEG by a factor of
$0.849\pm 0.018$, determined from comparing the fluxes of both instruments in
the 4.5--6.0 keV band. This renormalisation incorporates both calibration
uncertainties as well as possible small differences in the spectra because of
slightly different time intervals (Table~\ref{tab:obslog}).

Swift UVOT data, taken simultaneously with the first Chandra/NuSTAR observation,
were extracted as described in \citet{mehdipour2017}. Lacking UVOT data for the
second observation, we used the fluxes from the first observation also for the
second observation. Due to the short exposure time compared to the other X-ray
instruments we did not use the Swift XRT data for our spectral analysis.

Archival ASCA spectra were taken from the Tartarus database \citep{turner2001}.
This database also contains the response matrices and background spectra. ASCA
observed NGC~3783 twice in 1993, four times in 1996 and three times in 2000. The
observations in each year were taken within the same week and we have combined
the spectra into a single spectrum for each year. Although there are moderate
differences between the spectra within one week, the differences between the
average 1993 and 1996 spectra are much larger \citep{george1998}. In this paper
we focus on these time-averaged ASCA spectra.

We also combined SIS0 and SIS1 data. This data addition was performed using the
\textit{Heasoft addspec} utility. The 1993 and 1996 spectra show a strong
spectral difference at low energies \citep{george1998}. A plot of all three
spectra shows that the spectrum in 2000 is very similar to the 1996 spectrum,
except for a $\sim 30$\% weaker soft excess at 0.5~keV (Fig.~\ref{fig:asca}).
Given their higher spectral resolution, we use here data taken with SIS0 and
SIS1 detectors and ignore the GIS2 and GIS3 data.

Lacking contemporaneous UV data for the ASCA epochs, we have assumed that the
UVOT fluxes as observed in August 2016 were the same as in 1993 and 1996. While
in reality there may be a significant difference, this choice prevents that the
Comptonised soft excess component becomes unrealistically unconstrained due to
extrapolation from the X-ray band. Therefore we fit the ASCA spectra together
with the UVOT data, but we only report the goodness of fit for the ASCA part of
the data. 

The properties of the obscurer do not depend significantly on the precise UV
flux level. For instance, lowering the UV flux by a factor of two for the 1993
ASCA data affects the parameters of the comptonised soft excess, but the column
density and covering factor of the obscurer remain unchanged within their
statistical uncertainties.

\subsection{Spectral modelling}

\begin{table*}[!htbp]
\caption{Best-fit parameters derived in this work for the August 2016
spectra of NGC~3783 and the ASCA spectra taken in 1993, 1996 and 2000.
For comparison we also list the best-fit parameters
as obtained by \citet{mehdipour2017} for the December 2016 spectra.}
\label{tab:fits}
\centerline{
\begin{tabular}{lccccccc}
\hline\hline
Date & 12-1993 & 7-1996 & 1-2000 & 22-8-2016 & 25-8-2016 & 11-12-2016 & 21-12-2016 \\
\hline\noalign{\smallskip}
\multicolumn{8}{l}{Component 1} \\
$N_{\rm H}$ ($10^{27}$~m$^{-2}$) & --- & --- & --- & --- & --- & $1.5\pm 0.2$ & $2.0\pm 0.2$ \\
$C_f$                           & --- & --- & --- & --- & --- & $0.47\pm 0.10$ & $0.38\pm 0.03$ \\
\hline\noalign{\smallskip}
\multicolumn{8}{l}{Component 2} \\
$N_{\rm H}$ ($10^{27}$~m$^{-2}$) & $0.17\pm 0.02$ & $0.11\pm 0.02$ & $0.036\pm 0.003$ & $0.18\pm 0.02$ & $0.44\pm 0.06$ & $0.8\pm 0.2$ & $0.3\pm 0.1$ \\
$C_f$                            & $0.69\pm 0.02$ & $0.27\pm 0.03$ & $1.00\pm 0.05$   & $0.63\pm 0.05$ & $0.45\pm 0.02$ & $0.51\pm 0.10$ & $0.48\pm 0.03$ \\
\hline\noalign{\smallskip}
\multicolumn{8}{l}{Power law} \\
$\Gamma$ & 1.91 & 1.78 & 1.86 & 1.81 & 1.78 & 1.71 & 1.75 \\
$L_{2-10\,{\rm keV}}$ ($10^{36}$~W) & 2.56 & 3.17 & 3.21 & 1.31 & 1.10 & 0.95 & 1.16 \\
\hline\noalign{\smallskip}
\multicolumn{8}{l}{Comptonised disk} \\
$L$ ($10^{36}$~W) & 20.3 & 8.7 & 7.6 & 11.8 & 8.0  & 15.7 & 17.7\\
$T_0$ (eV) & 2.3 & 1.3 & 1.5 & 1.5  & 1.3  & 1.1  & 1.1 \\
$T_1$ (eV) & 142 & 141 & 132 & 145  & 129  & 128  & 131 \\
\hline\noalign{\smallskip}
\multicolumn{8}{l}{Goodness of fit} \\
C-stat          & 539 & 520 & 283 & 5077 & 4721 & 2288 & 2285 \\
Expected C-stat & 228 & 184 & 140 & 4706 & 4497 & 1539 & 1542 \\
\hline\noalign{\smallskip}
\end{tabular}
}
\end{table*}

All spectral modelling was done using the SPEX package \citep{kaastra1996}. The
basic spectral model is summarised here in some detail. It is the model used by
\citet{mehdipour2017} and that paper justifies the choice for the model and the
adopted parameters.

The spectral model is described by the following
components (in \textsl{slanted} font the SPEX names of the components):
\newline\noindent 1. Power law (\textsl{pow}) with exponential cut-off
(\textsl{etau}) at high energies (340 keV) and low energies (1.53 eV)
representing the dominant emission component.
\newline\noindent 2. Comptonised soft X-ray excess (\textsl{comt}).
\newline\noindent 3. Eight X-ray broad lines, modeled as delta lines with
Gaussian velocity broadening (\textsl{delt} with \textsl{vgau} applied), for
\ion{O}{viii} and \ion{N}{vii} Ly$\alpha$ and the He-like triplets of
\ion{O}{vii} and \ion{N}{vi}.

These emission components 1--3 are then fed successively through the obscurer
(modelled by two \textsl{xabs} components and one \textsl{pion} component for
the high ionisation component) and the nine warm absorber components (all
modelled with the \textsl{pion} component, accounting only for their absorption
properties). Note that for each of these succesive absorption steps the ionising
spectrum becomes weaker. The finally transmitted spectrum is then cosmologically
redshifted and corrected for absorption in our Galaxy. This Galactic absorption
is done using the \textsl{hot} model for the X-ray part of the spectrum ($E>
13.6$~eV) and the \textsl{ebv} model for the optical/UV part. Technically, this
is achieved by duplicating these components and by deleting either the UV part
or the X-ray part with an \textsl{etau} component with large, constant optical
depth over the relevant part of the spectrum.

Further we have three other emission components:
\newline\noindent 4. Power law (\textsl{pow}) with exponential cut-off (\textsl{etau}) 
at high energies (340 keV) and low energies (1.2 eV)
\newline\noindent 5. Comptonised soft excess (\textsl{comt})
\newline\noindent 6. Disc reflection component (\textsl{refl}) plus a narrow Gaussian line
(to represent the Fe K$\beta$ line which is not included in the \textsl{refl} model) 
from distant reflecting material.

These three emission components 4--6 represent the long-term time-average ionising
continuum for the distant X-ray narrow line emission region. They are fed through the
two \textsl{pion} emission components representing the narrow line emission region,
and then we delete the flux of components 4--5 by applying again an
\textsl{etau} component with high optical depth, in order not to duplicate the
actually seen continuum already accounted for by components 1 and 2. The reflection
component is new here and accounted for only in this part of the calculation.

Finally, we add two constant emission terms:
\newline\noindent 7. The host galaxy optical/UV continuum in the spectral extraction region.
\newline\noindent 8. The optical emission line and continuum spectral features.

Both of these last two components are represented with a \textsl{file}
component, and they are described in more detail by \citet{mehdipour2017}.

The flux of NGC~3783 in the soft band was relatively low during these
observations, and the effective area of the HETGS below 1~keV has decreased
significantly over its lifetime. Therefore the signal to noise ratio of the
HETGS data at lower energies is not optimal. This means we have to constrain the
number of free parameters.

We use as the baseline the model parameters for the obscured spectrum
as taken on December 11, 2016 by XMM-Newton, NuSTAR and HST-COS. This model has
been described in detail by \citet{mehdipour2017}. Basically,
\citet{mehdipour2017} first determine the properties of the outflow ("warm
absorber") from archival XMM-Newton and Chandra HETGS spectra obtained between
2000--2001 and in 2013. The outflow has 9 components, all modeled with the
\textsl{pion} model of SPEX, and with different combinations of ionisation
parameter and outflow velocity. For the obscured state observed in December 2016
they then assume that all parameters of the outflow components remain the same
except for the ionisation parameters, which are calculated (not fitted) from the
ionising continuum that the outflow receives assuming that the gas density and
distance to the ionising source have remained constant between 2001 and 2016.
This ionising continuum is determined from fitting the observed spectrum with
the primary continuum parameters (power law, reflection component and
comptonised soft excess) and intervening obscurer parameters left free and
including the outflow as described above.

In addition to the primary continuum, affected by both the obscuring medium and
the warm absorber, the \citet{mehdipour2017} model also contains some emission
from gas far away from the nucleus which produces narrow emission features.

\begin{figure}[!ht]
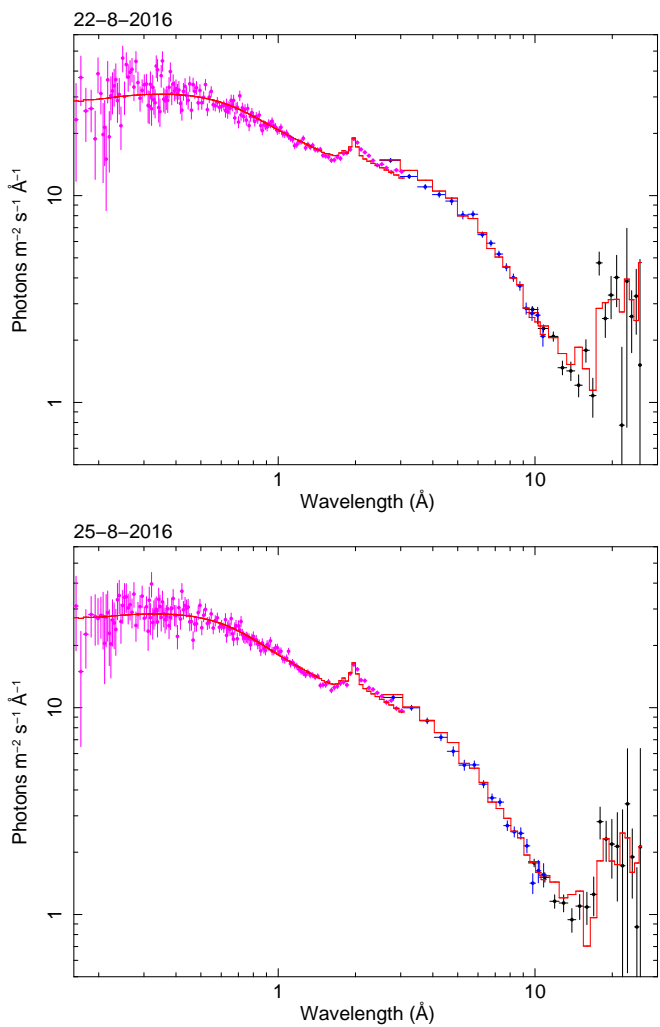

\resizebox{\hsize}{!}{\includegraphics[angle=-90]{fit09broad.ps}}
\resizebox{\hsize}{!}{\includegraphics[angle=-90]{fit10broad.ps}}
\caption{Fluxed spectra of NGC~3783 on August 22 and August 25, 2016 using the
model described in the text. Red curve: best-fit model; purple crosses: NuSTAR
data; blue and black crosses: HEG and MEG data, both rebinned by a factor of 100
for display purposes.}
\label{fig:spectra_broad}
\end{figure}

\begin{figure*}[!ht]
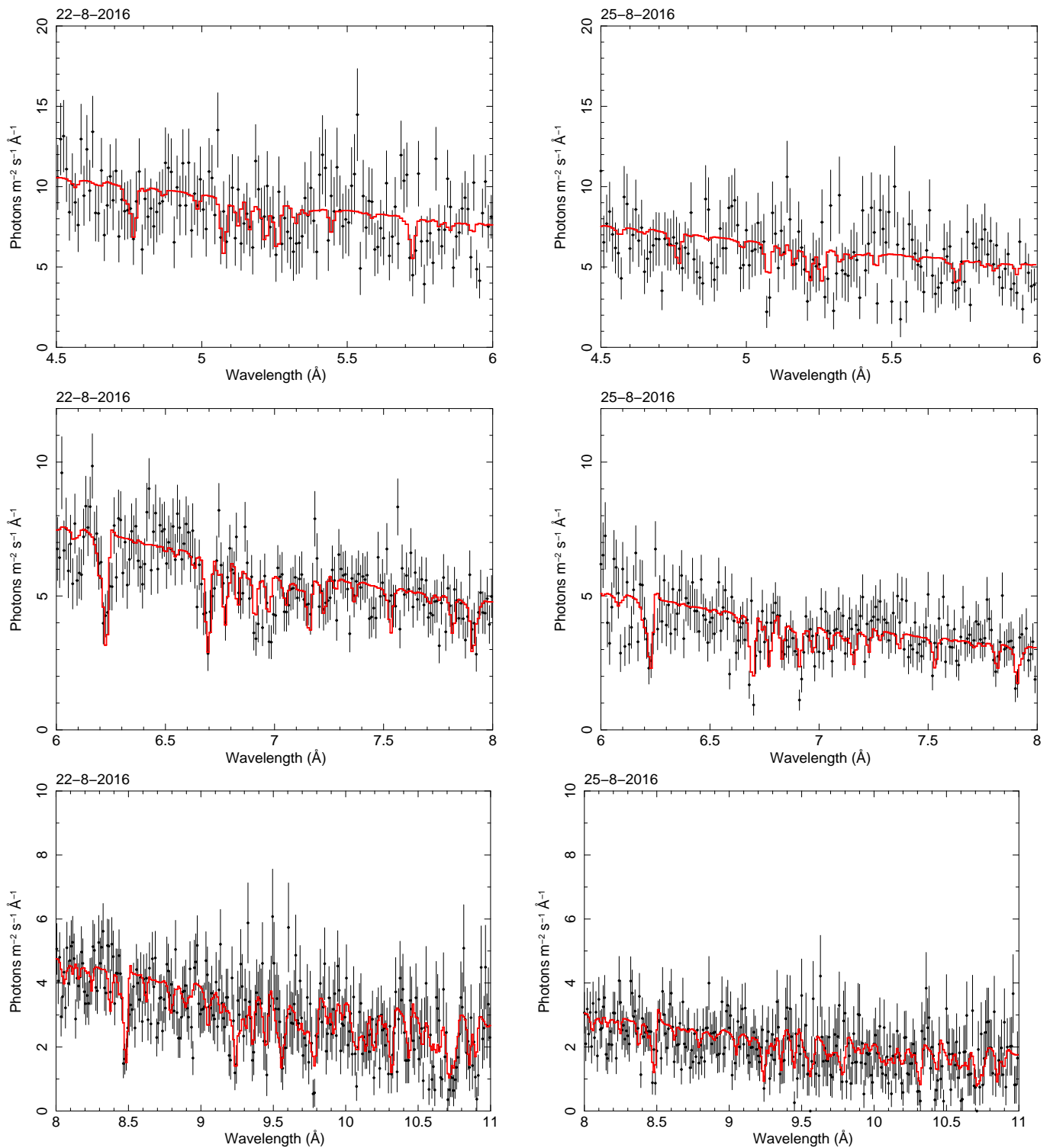

\resizebox{0.49\hsize}{!}{\includegraphics[angle=-90]{fit09fullb1.ps}}
\resizebox{0.49\hsize}{!}{\includegraphics[angle=-90]{fit10fullb1.ps}}
\resizebox{0.49\hsize}{!}{\includegraphics[angle=-90]{fit09fullb2.ps}}
\resizebox{0.49\hsize}{!}{\includegraphics[angle=-90]{fit10fullb2.ps}}
\resizebox{0.49\hsize}{!}{\includegraphics[angle=-90]{fit09fullb3.ps}}
\resizebox{0.49\hsize}{!}{\includegraphics[angle=-90]{fit10fullb3.ps}}
\caption{Fluxed spectra of NGC~3783 on August 22, 2016 (left)
and August 25, 2016 (right) using the
model described in the text. Red curve: best-fit model; 
black crosses: MEG data.}
\label{fig:spectra_full0910}
\end{figure*}

\begin{figure}[!ht]
\resizebox{\hsize}{!}{\includegraphics[angle=-90]{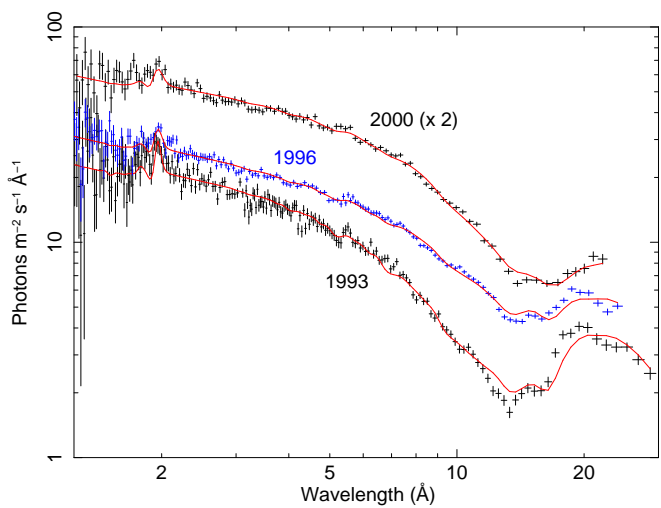}}
\caption{ASCA SIS spectra taken in 1993 (black crosses, lower spectrum), 1996
(blue crosses, middle spectrum) and 2000 (black crosses, upper spectrum, for
display purposes scaled by a factor of two), together with the best-fit models
as shown by the red lines.}
\label{fig:asca}
\end{figure}

In our model of the present August 2016 data we freeze the parameters of these
narrow emission components to the ones found in the December 2016 spectra. The
present HETGS data are not extremely sensitive to these emission features, so
this choice does not affect our final model. Again related to sensitivity
issues, we keep the reflection component constant to the December values, and
also the optical depth of the comptonised soft excess is kept constant. The only
free parameters of our model are therefore the normalisation, photon index and
exponential cut-off of the primary power law component, the normalisation, seed
photon temperature $T_0$ and electron temperature $T_1$ of the comptonised soft
excess, and the column densities and covering factors of the obscuring gas. All
other parameters are frozen to the December 2016 values or calculated assuming
that $n r^2 = L / \xi$ is constant for each component, with $n$, $r$, $L$ and
$\xi$ the hydrogen density, distance to the ionising source, ionising luminosity
and ionisation parameter, respectively.

For the December 2016 spectrum, the presence or absence of specific ions
in the high-resolution UV spectra taken simultaneously with HST/COS allowed to
constrain the ionisation parameter of the obscurer. In the present case, there
are no simultaneous high-resolution UV spectra. Therefore, for the obscurer we
adopt the same ionisation parameters as in December 2016, which is also
justified because the X-ray transmission of the obscurer is not very sensitive
to this ionisation parameter.

All fits were done using the full set of X-ray and UV spectra. Because
we lack simultaneous UV data for all but one epoch, the UV spectra merely serve
to constrain the spectral model within realistic limits, as discussed before.
Due to UV variability (see for instance Fig.~\ref{fig:swift}) the actual UV flux
may differ from our adopted values. We have tested that a factor of two higher
or lower UV flux does not affect the derived fit parameters as listed in
Table~\ref{tab:fits} in a strong way. In general, the UV flux points match the
best-fit models within 20\%.

For the spectral fitting, we use the C-statistics as implemented in SPEX
\citep{kaastra2017}. The best-fit parameters are shown in Table~\ref{tab:fits}.
We find that one obscuring component is sufficient to model all these data.
Fig.~\ref{fig:spectra_broad} shows how well our model matches the broad-band
spectrum in August 2016, and Fig.~\ref{fig:spectra_full0910} shows the
corresponding Chandra spectra in greater detail in the 4.5--11~\AA\ band, where
the signal is strong enough to reveal individual absorption lines.
Fig.~\ref{fig:asca} shows the best fit for the ASCA data. 

\subsection{Swift lightcurves}

\begin{figure*}[!ht]
\resizebox{\hsize}{!}{\includegraphics[angle=0]{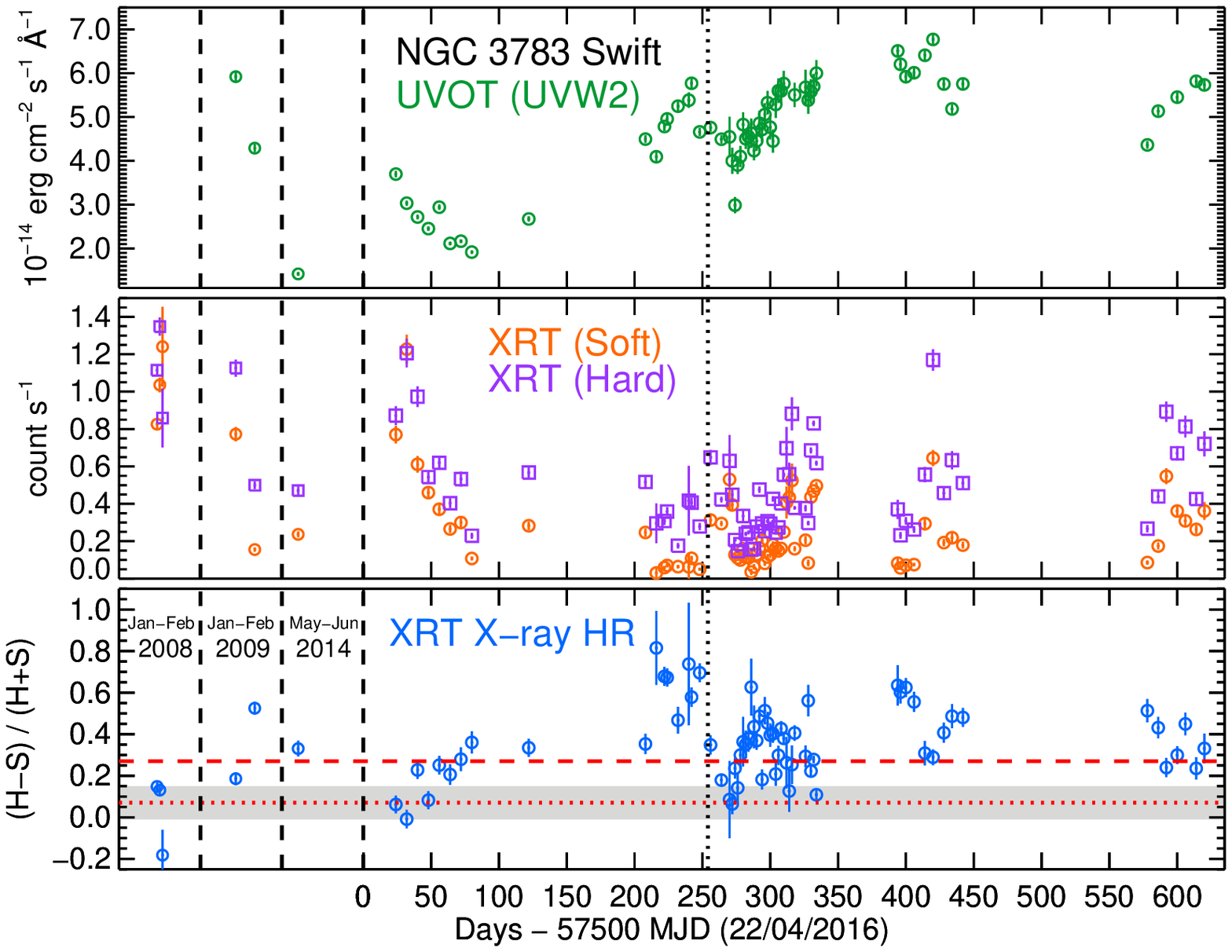}} 
\caption{Swift lightcurves in different bands. The vertical dotted line
indicates 1 January 2017. The data have been averaged over two days. 
The last datapoint was taken on January 3, 2018.
Top panel:
UVOT flux; middle panel: Soft (0.3--1.5 keV) and hard (1.5--10 keV) X-ray count
rate; bottom panel: hardness ratio $R=(H-S)/(H+S)$ with $S$ and $H$ the soft and
hard X-ray count rates. Dotted horizontal line: average hardness without
obscuration as described in the text; the gray area corresponds with its typical
variations due to variations in the continuum. Dashed horizontal line: hardness
above which there is strong obscuration (see text).}
\label{fig:swift}
\end{figure*}

Fig.~\ref{fig:swift} shows the Swift lightcurves in different bands as well as
the X-ray hardness ratio $R$, which provides an indication for obscuration when
it is high enough. The hardness is defined by $R=(H-S)/(H+S)$ with $S$ and $H$
the soft and hard X-ray count rates in the 0.3--1.5 and 1.5--10~keV bands,
respectively. The model of \citet{mehdipour2017} for the unobscured spectrum
based on all archival Chandra HETGS and XMM-Newton data taken in 2000 and 2001
yields a predicted hardness ratio $R=0.07$. In that model, the photon index of
the primary power law is 1.60. Making the spectrum harder, with a photon index
of 1.40, increases $R$ to 0.15. Alternatively, removing completely the soft
excess also raises $R$ to a value of 0.15. We therefore adopt here a range of
$-0.01$ to $+0.15$ as the fiducial hardness range corresponding to the natural
variations of the primary AGN continuum.

\begin{figure}[!ht]
\resizebox{\hsize}{!}{\includegraphics[angle=-90]{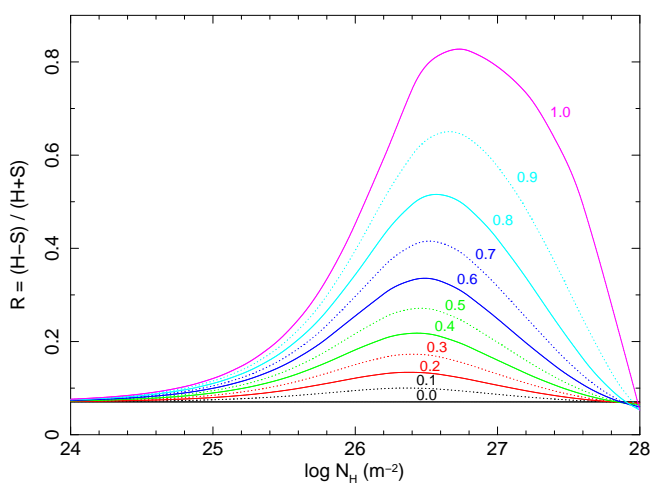}} 
\caption{Predicted Swift hardness ratio $R$ as a function of the obscuring column
density $N_{\rm H}$ (curves) and the covering factor $C_f$ (labels).}
\label{fig:predhard}
\end{figure}

Using the same parameters for the unobscured spectrum based on the 2000 and 2001
spectra, we predict the hardness $R$ in case there is additional obscuration
like we modeled for the August 2016 Chandra spectra and the ASCA spectra. Our
results are shown in Fig.~\ref{fig:predhard}. For zero covering factor, the
hardness ratio is 0.07. For larger covering factors, $R$ increases with
increasing column density. It reaches a maximum for column densities between
3--5$\times 10^{26}$~m$^{-2}$ and then declines. At high column densities, the
full covered primary continuum below 10~keV gets blocked. In that case only the
distant, uncovered and constant reflection component of our model remains
present, plus the uncovered fraction of the primary continuum. This explains the
decline of the hardness ratio at high column densities.

The maximum value that $R$ can achieve is about 0.8 for full covering and a
column density of about $5\times 10^{26}$~m$^{-2}$. Indeed Fig.~\ref{fig:swift}
shows no epochs with $R$ above this value.

In Fig.~\ref{fig:swift} we have indicated the level of $R=0.27$ by a dashed
horizontal line. This value corresponds to the maximum hardness for covering
factor 0.5. Values $R>0.27$ correspond to covering factors of at least 50\% and
column densities larger than $5\times 10^{25}$~m$^{-2}$). We may call these
strongly obscured states.

We note that the flux decrease due to the obscuration is enhanced by the distant
warm absorber: it receives less ionising radiation, recombines and gets higher
opacity. Further, the model for $R$ presented in Fig.~\ref{fig:predhard} only
holds for a single obscuration component. In case of multiple components, like
in December 2016, the situation is more complex because there are more free
parameters. But in that case the flux decrease is even stronger due to the
combined effect of all components.

There are several epochs where the hardness is large and strong obscuration is
present. One event occurred in 2009, but with the sparse sampling in that year
it is hard to put that into context. The second event is seen in December 2016
(days $\sim$210--250 in Fig.~\ref{fig:swift}) and is discussed in detail by
\citet{mehdipour2017}. From mid January 2017 (day 270) the hardness increased
gradually over two weeks time to reach a similar level as in December 2016. It
remained high for about two months, with quite some variations. The latest data
in Fall 2017, taken after a data gap in Summer, show again such a hard state.

Interestingly, in August 2016 when the Chandra data were taken (day 122 in
Fig.~\ref{fig:swift}), the Swift hardness ratio was also high (0.35) but below
the trigger level for the XMM-Newton program. As we have shown, in this state
there is also obscuration but less than in December 2016. The Swift hardness
ratio curve shows several other epochs with similar hardness, like in 2014 and
March 2016. In fact, 66~\% of all Swift data points show a hardness larger than
0.27, although we must take care because the spacing of the data points is not
even.

\section{Discussion}

\subsection{Obscuration in August 2016}

The spectral fits that we have obtained can be characterised by the deep broad
dip visible between 15--17~\AA\ (Fig.~\ref{fig:spectra_broad}). This is caused
by a deep (optical depth 1.7) \ion{O}{vii} edge produced by the obscuring
material, and enhanced by a multitude of iron lines from the warm absorber. Such
a deep edge is not present in the normal, unobscured spectrum of NGC~3783.

Our fit reproduces the main features of the present NGC~3783 spectra well, as
can be seen from both the rebinned broad-band spectra
(Fig.~\ref{fig:spectra_broad}) and the detailed HETGS spectra showing good
matches of the strongest absorption lines with the model
(Fig.~\ref{fig:spectra_full0910}).

The ionising continuum in August 2016 is slightly different from the continuum
four months later. While the luminosity of the power law component is similar
for both epochs, the comptonised soft excess is about 40\% weaker in August,
which can also be seen more directly from the lower flux observed with the Swift
UVOT.

\begin{figure}[!ht]
\resizebox{\hsize}{!}{\includegraphics[angle=-90]{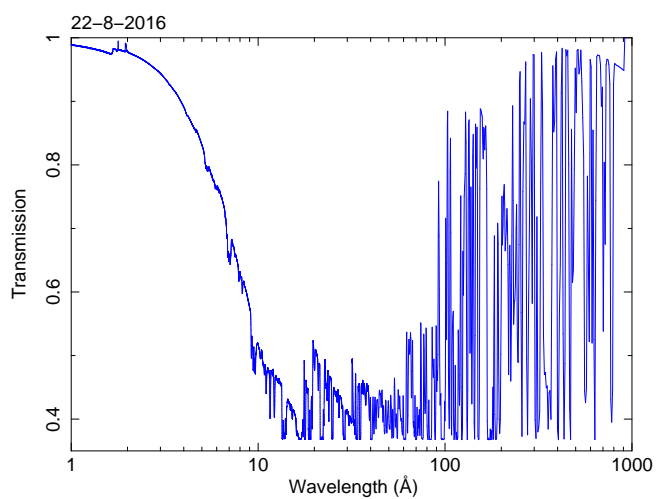}}
\caption{Transmission of the obscurer on 22 August 2016.}
\label{fig:modrat}
\end{figure}

Interestingly, NGC~3783 shows clear evidence for obscuration in August 2016,
although lower than four months later in December. While the covering factor of
the obscurer is of the same order of magnitude or even slightly larger than in
December, the column density is an order of magnitude lower, but still
sufficient to remove a significant fraction of the X-ray flux. We show the
transmission profile of the obscurer in Fig.~\ref{fig:modrat}. Because of the
partial covering, 37\% of the X-ray flux is always transmitted. It is seen from
this figure that the transmission curve shows a lot of structure. At the
shortest wavelengths, the transmission drops gradually from unity to close to
the residual level of 37\%, with a few broadened absorption lines superimposed.
In the 15--100~\AA\ band, there is detailed structure with a combination of deep
edges from oxygen ions and strong absorption lines. But also the opacity in the
EUV band up to the Lyman edge is structured, and a significant fraction of the
ionising EUV continuum is absorbed.

It should be noted that, like in other cases, the obscuring material has a
strong recombining effect on the warm absorber. The total X-ray blocking from
the combined obscurer and recombined warm absorber is much stronger than during
unobscured phases like seen in 2000 and 2001.

Our model predicts that the 2s--2p transitions of \ion{C}{iv} lines at 1548 and
1550~\AA, the \ion{N}{v} 2s--2p transitions at 1238 and 1242~\AA\ and the
\ion{H}{i} Ly$\alpha$ line from the obscuring medium all have optical depths of
order unity, so should produce detectable UV absorption lines. On the other
hand, the optical depth in the \ion{O}{vi} 2s--2p lines is of order 100. This
shows that the predicted optical depth of the UV lines of the obscurer is very
sensitive to the parameters of this obscuring medium. Unfortunately, in August
2016 no UV spectra of NGC~3783 were taken that could validate our model.

\subsection{Obscuration at other epochs}

We have also investigated possible obscuration using observations from other
satellites. The spectra taken by BeppoSAX in June 1998 \citep{derosa2002} and
Suzaku in July 2009 \citep{reis2012} do not show apparent evidence for
obscuration. However the ASCA spectra taken in 1993, when compared to later data
taken in 1996, show evidence for strong spectral differences in the soft X-ray
band \citep{george1998}. These were interpreted by the authors as corresponding
to absorbing gas with more than one component, where one absorbing component
with a typical column density of $10^{25}$~m$^{-2}$ that was present in 1993
would have moved out of the line of sight in 1996. While the spectral resolution
of the ASCA SIS detectors in the soft X-ray band hampers a detailed study like
is possible with the XMM-Newton data, the basic conclusions of
\citet{george1998} are consistent with a scenario where in 1993 moderate
obscuration is present.

This is confirmed by our own fits to these data. The number of free parameters
of our model is very small: only the power law and comptonised soft component
parameters, the column density and covering factor of the obscurer are left free
in the fit. The warm absorber is completely determined by its column densities
and ionisation state as derived from the grating spectra taken in 2000 and 2001,
adjusted for the contemporaneous ionising SED. The column density and covering
factor we derive for the 1993 spectrum are remarkably similar to those on August
22, 2016, with a two times higher continuum flux in 1993.

While our model for the ASCA data matches the broad-band dip in the spectrum
well, the fit is not perfect and shows a few systematic deviations at the
10--20\% level near 0.65~keV (20~\AA) and 1~keV. These might be caused by a
combination of various processes, like our constraint on the UV flux, secular
changes of some components over the 23 years between 1993 and 2016, calibration
issues etc. The relatively low spectral resolution of ASCA in the important soft
X-ray band prohibits perhaps to answer this uniquely. However, it is surprising
that with so few free parameters we still can get a rough fit that catches the
main obscuration trough. In any case, both our analysis and the original
modeling by \citet{george1998} indicate the presence of additional obscuring
material in the 1993 ASCA spectrum.

Finally we note that the spectrum taken in 2000 has the lowest column density of
all three ASCA spectra, with only $3.6\times 10^{25}$~m$^{-2}$
(Table~\ref{tab:fits}). This is only 20\% of the column density of the distant
warm absorber for the same ionisation parameter, and therefore, despite its
nominal statistical uncertainty, we do not claim that there is obscuration in
this spectrum. The systematic uncertainties are likely the largest for this
spectrum, with the highest ratio of the 2--10 keV power law flux to the
comptonised component flux of all spectra (see Table~\ref{tab:fits}). It is also
consistent with our assumption that the XMM-Newton and Chandra spectra taken in
2000 and 2001, the same epoch as this ASCA spectrum, were free of obscuration.  

\subsection{The frequency of obscuration}

Obscuration appears to be more frequent in NGC~3783 than previously thought.
About half of all Swift observations show evidence for moderate to strong
obscuration, and also in the more distant past, obscuration events have occurred
like those seen with ASCA in 1993 and the three RXTE events mentioned in the
introduction, see \citet{markowitz2014}. 

In fact, the sparsity of most data makes it hard to define what an obscuration
event is. The densely sampled part of the Swift lightcurve in Spring 2017 shows
strong variability, with almost always some level of obscuration present. It is
a matter of semantics whether one calls the peaks of this obscuration events, or
whether one calls the entire few months period an obscuration event.

Interestingly, our modeling indicates that even the ASCA 1996 spectrum may show
some obscuration, albeit with a low column density of $\sim 10^{26}$~m$^{-2}$
and only 27\% covering factor. Such values may also occur during the "softest"
intervals of longer-lasting obscuration periods as seen by Swift. In most other
spectral analyses, such a parameter range of the obscuring gas may have been
easily overlooked. The continuum obscuration may be compensated by modifying the
shape or relative intensity of other spectral parameters like the main power
law, the soft excess or the reflection component. 

In the cases that we studied, obscuration could be proven because of the
availability of high-resolution X-ray spectra combined with broad-band SED
information. Whenever also high-resolution UV spectra are available, like for
NGC~5548 \citep{kaastra2014s} or the December 2016 observations of NGC~3783
\citep{mehdipour2017}, the dynamical state of the obscuring stream can be
determined from the UV line profiles. Also, the observed UV lines set bounds on
the allowed range of ionisation parameter for the obscuring gas, something that
cannot be done in the X-rays in these cases due to the lack of spectral features.

The higher frequency of obscuration events that we find here also implies that
for the distant emitting gas of the ionisation cones or narrow line region, the
effective ionising continuum radiation, as averaged over several years for gas
at pc distance scales, may be smaller than if one would adopt the
contemporaneous SED of an unobscured epoch. This will affect the modelling of
the photoionised gas in those regions.

Finally, as can be deduced from the light curves, the obscuration is highly
variable in terms of column density and covering factor. Similar conclusions
have been reached for the obscuration in NGC~5548. The obscuring streams
apparently have relatively large density and therefore short recombination
times, hence adjust their ionisation state quickly to changes of the ionising
continuum. The strong variability may also be caused by a rather inhomogeneous
and irregular matter distribution within the outflow, causing rapid column
density and covering factor variations.

\subsection{Future prospects}

\begin{figure}[!ht]
\resizebox{\hsize}{!}{\includegraphics[angle=-90]{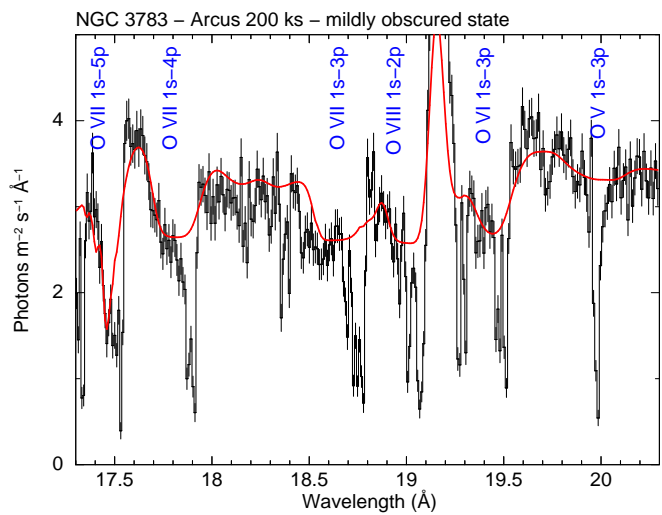}}
\caption{Part of a simulated spectrum for the Arcus mission for NGC~3783 with
the same source parameters as it showed on 22 August 2016 (this paper). The
simulated spectrum is shown as the black histogram with error bars. The red
curve shows the obscured continuum that the warm absorber receives, scaled down
by a factor of 3.3. The narrow \ion{O}{viii} Lyman alpha emission line at
19.15~\AA\ is from a remote region and is cut-off in this plot. This figure
shows that Arcus will be able to separate the narrow absorption lines of the
warm absorber from the broad absorption lines from the obscurer.}
\label{fig:arcus}
\end{figure}

To study obscuration effects to their full extend and in a larger sample
requires monitoring with multiple instruments. At this moment, we need X-ray
grating spectra to characterise the X-ray obscuration, and in addition
high-resolution UV spectra to confirm \citep[like in NGC~5548,][]{kaastra2014s}
or measure \citep[like in NGC~3783,][]{mehdipour2017} the ionisation parameter
of the outflow. UV spectra are also needed to measure the outflow velocity of
the obscurer. Moreover, because the obscuration is highly variable and
transient, it requires either extensive monitoring or triggered observations.

Proposed new X-ray instrumentation may offer another solution. The Arcus mission
\citep{smith2016} has an order of magnitude higher effective area and spectral
resolution compared with the currently most sensitive spectrometer, the RGS on
board of XMM-Newton. This is sufficient to separate the narrow absorption lines
of the warm absorber from the broader absorption lines of the obscurer in the
X-ray band, which is impossible with our present instrumentation. See
Fig.~\ref{fig:arcus} for a simulation of the August 22, 2016 spectrum of
NGC~3783. This does not imply that we do not need high-resolution UV spectra in
addition. The combination of high-resolution X-ray and UV spectra would allow to
find the covering factors in both bands individually. This allows to much better
constrain the geometry of the inner parts of the AGN.

\section{Conclusions}

By analysing archival Chandra HETGS spectra of NGC~3783, we have found that also
in August 2016 obscuration of the nucleus took place, albeit with a lower column
density than four months later in December 2016. We have also identified an
epoch when ASCA is likely to have caught the source in a mildly obscured state.
Swift monitoring indicates that obscuration can be seen in a significant
fraction of all observations. Obscuring column densities $>5 \times
10^{25}$~m$^{-2}$ occur in more than 66\% of all Swift observations. Obscuration
of the nuclear radiation by gas at broad-line distances may be more common than
we thought, at least in NGC~3783. Whether this holds for other AGN as well
requires more study.

\begin{acknowledgements}

SRON is supported financially by NWO, the Netherlands Organization for
Scientific Research. The research at the Technion is supported by the I-CORE
program of the Planning and Budgeting Committee (grant number 1937/12). EB
acknowledges funding from the European Union's Horizon 2020 research and
innovation programme under the Marie Sklodowska-Curie grant agreement no.
655324. SB and MC acknowledge financial support from the Italian Space
Agency under grant ASI-INAF I/037/12/0, and under the agreement ASI-INAF
n.2017-14-H.O. BDM acknowledges support from Polish National Science Centre
grant Polonez 2016/21/P/ST9/04025. POP aknowledges support from the CNES and
French PNHE.

\end{acknowledgements}

\bibliographystyle{aa}
\bibliography{paper}

\end{document}